\documentclass[11pt,a4paper]{article}%
\usepackage{jheppub}
\usepackage{amsmath}
\usepackage{amssymb}
\usepackage{graphicx}
\usepackage{epstopdf}
\usepackage{amsfonts}
\usepackage{amssymb}
\usepackage{booktabs, color, cancel, mathrsfs, enumerate, setspace, svn}
\providecommand{\U}[1]{\protect\rule{.1in}{.1in}}
\affiliation[a]{Center for Theoretical Physics and College of Physics, Jilin University, Changchun 130012, People's Republic of China}
\affiliation[b]{Max Planck Institute for Gravitational Physics (Albert Einstein Institute),
	Am M\"uhlenberg 1, 14476 Golm, Germany}
\affiliation[c]{Department of Electrophysics, National Chiao Tung University, Hsinchu, ROC}
\affiliation[d]{Physics Division, National Center for Theoretical Sciences, Hsinchu, ROC}
\affiliation[e]{School of physics, University of Chinese Academy of Sciences, Beijing 100049, China}
\affiliation[f]{Kavli Insititute for Theoretical Sciences, University of Chinese Academy of Sciences, Beijing 100049, China}
\emailAdd{hesong@jlu.edu.cn}
\emailAdd{yiyang@mail.nctu.edu.tw}
\emailAdd{phy.pro.phy@gmail.com}
\abstract{We explore the effect of the magnetic field on the QCD phase transition through AdS/CFT correspondence. By introducing an anisotropic magnetic field in the Einstein-Maxwell-Scalar system, a family of analytic solutions is obtained by the potential reconstruction method where the contribution of the magnetic field in the blackening background can be analytically derived. After imposing the kinetic gauge function by requesting the linear Regge spectrum of $J/\psi$ mesons, the contribution of the magnetic field phase diagram can be demonstrated. The results show that the transition temperature will be raising as the magnetic field increases, which is the so-call magnetic catalysis effect. However, if the system is in a strong enough magnetic environment, the transition temperature will be cool down and display the inverse catalysis effect.}
\begin{document}
	\title{Analytic Study of Magnetic Catalysis in Holographic QCD}
	\author{Song He${^{a,b}}$, Yi Yang${^{c,d}}$, Pei-Hung Yuan${^{e,f}}$}
	\maketitle

%

\setcounter{equation}{0}
\renewcommand{\theequation}{\arabic{section}.\arabic{equation}}%

\section{Introduction}
The recent experiments of heavy ion collision at the Relativistic Heavy Ion Collider (RHIC) and the Large Hadron Collider (LHC) have produced a strongly coupled quark-gluon plasma (QGP) and explored the QCD phase structure under finite baryon density and strong magnetic fields \cite{0907.1396,1103.4239,1111.1949,1201.5108}. In addition, strong magnetic fields are present in neutron stars and magnetars \cite{Astrophys.J.392(1992)L9,0804.0250}, as well as in the cosmological phase transition during the early stages of the universe \cite{9308270}. Therefore, understanding the effect of magnetic field on QCD phase transition is an important task in fundamental physics.

The early investigation by lattice QCD showed that the phase transition temperature increases with the external magnetic field, i.e. magnetic catalysis (MC) \cite{1005.5365,1203.3360,1209.0374}. The MC phenomenon has been confirmed by many effective QCD theories \cite{9405262,9412257,9509320,9905116,0010211,0205348,0701090,0810.3693,1003.0047,1004.2712,1012.1291,1104.5167,1107.4737,1112.5137,1203.4330,1207.5081,1208.0917,1305.4751,1311.3964,1312.6733} as well as holographic QCD (hQCD) models \cite{1110.5902,1511.04042}. See \cite{1207.5081} for a review. However, later lattice simulations revealed the opposite results by considering the physics quark mass that the phase transition temperature decreases with the external magnetic field, i.e. inverse magnetic catalysis (IMC) \cite{1103.2080,1111.4956,1206.4205,1303.1328,1303.3972,1310.7876,1312.5628,1504.08280,1607.08160,1711.02884,1909.09547}. The IMC phenomenon has been investigated in numerous literature \cite{0803.3156,1201.5881,1207.3190,1207.7094,1208.0536,1208.5025,1209.1319,1211.6245,1305.1100,1404.5577,1404.6969,1406.3885,1406.7408,1407.3503,1410.5247,1412.1647,1502.08011,1509.01181,1602.06448,1603.03847,1712.08384,1907.01852}. Theoretically, we still do not fully understand the response of the microscopic mechanism for the magnetic field effects. The one interpretation is that the effect of magnetic field can reduce the number of the effective dimensions of the quarks, which leads to the chiral condensation \cite{1207.5081,1211.6245}. Comparing with the heavy quark, increasing temperature, the light sea quarks can be excited easily which induces chiral condensation of the light quarks. Furthermore, it is believed that the corresponding effects are due to the competition between the direct valence effect and the indirect sea effect \cite{1103.2080,1303.3972,1808.07008,1904.10296}. The valence effect induces dynamical mass generation and always causes MC phenomenon. While including the back-reaction of the sea quarks on the gauge fields and the screening effect of the gluon interactions will suppress the phase transition and causes the IMC phenomenon. On the other hand, the recent lattice simulations have demonstrated that the magnetic effect depends on quark mass. For light quarks, the sea effect wins the competition and induces IMC phenomenon. While for heavy quarks, the valence effect becomes more important and the phase transition behaves as MC phenomenon \cite{1808.07008,1904.10296}.

Investigating the QCD phase transitions has been a long crusade to understand the fundamental physics and astrophysics for decades. Since most perturbative QCD calculations and effective models suggest MC, people believe that IMC is due to the strongly coupled dynamics near the phase transition. However, the standard perturbative method can not treat the issue well because of the strongly coupled region. In addition, the non-perturbative techniques such as lattice QCD are facing the sign problem for finite baryon densities. In order to reveal the whole picture in the "temperature-chemical potential" plane, the holographic framework \cite{9711200,9802109,9802150} is a good candidate to fulfill the job. In hQCD, phase transitions in the presence of external magnetic field has been previously studied for both confinement-deconfinement phase transition \cite{1501.03262,1505.07894,1511.02721,1604.07197,1709.09258,1710.07310,1811.11724} and chiral condensation \cite{1010.0444,1012.4785,1110.5902,1208.0536,1303.5674,1307.6498,1511.02721,1511.04042,1511.05293,1512.06493,1602.08994,1604.06307,1610.04618,1611.06339,1706.05977,1707.00872,1807.11822,1811.04117}.

One of the ingredients of MC/IMC phenomena is anisotropy. A series study in \cite{1611.06339,1707.00872,1708.05691,1811.11724} reveals that the isotropy could be broken not only by introducing a magnetic field but also a relevant or marginal operator. In this work, we study the magnetic effect on phase transition in QCD for heavy quarks by holographic correspondence. We are going to explore how the phase transition is affected by the external magnetic field, which plays the role of anisotropy. We study a hQCD model by considering an uniform but anisotropic external magnetic field in the 5-dimensional Einstein-Maxwell-Scalar (EMS) system, which is dual to a 4-dimensional QCD theory. This background is a magnetic generalization of the EMS frameworks considered in \cite{1201.0820,1301.0385,1406.1865,1506.05930,1703.09184,1705.07587,1812.09676}, in which a family of analytic solutions were obtained, and the phase diagram and the equations of states in QCD have been extensively studied. To ensure the stability of the gravitational background, we check the null energy condition (NEC), which induces a constraint for the profile of scalar field. To study heavy quarks, we fit our parameters by the Regge linear spectrum of $J/\psi$ mesons. By calculating the phase transition temperature for different chemical potentials under the varied magnetic field to obtain the phase structure of QCD. We find the MC/IMC phenomenon for the small/large external magnetic field. In addition, we locate the critical end point (CEP) for the QCD phase transition in $(T,\mu, B)$ 3-dimensional phase diagram. We find that the CEP moves to the lower chemical potential with a growing magnetic field. Furthermore, the CEP forms a closed boundary in the $(\mu, B)$ plane, beyond which the phase transition becomes crossover.

The organization of the remaining parts of this paper is as follows. The hQCD model with anisotropic constant magnetic field has been constructed and a class of analytic gravitational background solutions have been presented in Sec. 2. The holographic phase transition and CEP with magnetic effects have been investigated in Sec. 3. Finally, we devote to the conclusions and discussions. 



\section{Holographic Model}
Einstein-Maxwell-Scalar (EMS) system is one of the fundamental frameworks to build holographic models effectively describing the phenomena of the strongly coupled field theory. Using EMS system to construct hQCD models was advocated in \cite{1012.1864, 1108.2029}, and has been improved afterward in many literature. 

In this work, we will investigate the phase structure in the presence of external magnetic field in QCD analytically by using a well established analytical hQCD model in \cite{1301.0385,1406.1865,1506.05930,1703.09184,1705.07587,1812.09676}. Hopefully we can shed a light on this issue and deliver deeper insights for numerical simulations and experiments. 

\subsection{Holographic EMS System with Magnetic Field}
We consider a 5-dimensional EMS system as a thermal background for the corresponding QCD, which is constructed in Einstein frame as
\begin{eqnarray}
S_{B} &=& \frac{1}{16\pi G_{5}} \int d^{5}x\sqrt{-g}
\left[{R-\frac{f\left(\phi\right)}{4}F^{2}}
-\frac{1}{2} \left( \partial \phi \right)^2 
-V\left(\phi\right) \right], \label{eq_SB_Ef}
\end{eqnarray}where ${G}_{5}$ labels the 5-dimensional Newtonian constant, $R$ refers to the Riemann scalar, $\phi$ is a neutral scalar field, and ${{F}}_{\mu\nu} = \partial_{\mu}{A}_{\nu}-\partial_{\nu}{A}_{\mu}$ represents the field strength originated from the gauge field $A_\mu$. $f(\phi)$ is the gauge kinetic function and $V(\phi)$ is the potential of the $\phi$. The equations of motion are obtained by varying the Eq. (\ref{eq_SB_Ef})
\begin{eqnarray}
\nabla^{2}\phi &=& \frac{\partial V}{\partial\phi}+\frac{F^2}{4}\frac{\partial f}{\partial\phi}, \label{eq_eom_phi}\\
\nabla_{\mu}\left[ f(\phi)F^{\mu\nu} \right] &=&0, \label{eq_eom_A}\\
R_{\mu\nu}-\frac{1}{2} g_{\mu\nu}R &=& \frac{f(\phi)}{2} \left( F_{\mu\rho}F_{\nu}^{~\rho}-\frac{1}{4}g_{\mu\nu}F^{2}\right) +\frac{1}{2}\left[\partial_{\mu}\phi\partial_{\nu}\phi-\frac{1}{2}g_{\mu\nu}\left(\partial\phi\right)^{2}-g_{\mu\nu}V(\phi)\right] \label{eq_eom_g}.
\end{eqnarray}
To take account of external magnetic field, we consider the following anisotropic ansatz of the background blackening metric and ($\phi,~A_{\mu}$) fields in Einstein frame 
\begin{eqnarray}
ds^{2} &=& w_{E}(z)^2
\left[-b(z)dt^{2} +  g_{11}(z)  dx_1^{2} + g_{22}(z)\left(dx_2^{2} + dx_3^{2}\right) + \frac{dz^{2}}{b(z)}\right],\label{eq_metric}\\
\phi &=& \phi\left(z\right),~ A_{\mu}=\left(A_{t}(z),0,0,A_{3}(x_2),0\right), \label{eq_ansatz}
\end{eqnarray}
where $w_{E}(z) = \frac{e^{d(z)}}{z}$ is the warped factor with the sub-index $E$ labeling the Einstein frame, $d(z)$ is the deformed factor which describes the warping geometry deformed from $AdS$ spacetime, and $b(z)$ stands for the blackening factor which formulates the black hole background. Conventionally, $z = 0$ corresponds to the conformal boundary of the 5-dimensional space-time, and we have set the radial of $AdS_5$ to be unit by scale invariant.

We consider two components in the gauge field $A_\mu$ as Eq. (\ref{eq_ansatz}). The non-trivial temporal component $A_t(z)$ associates with the chemical potential $\mu$ and the spatial component $A_3=B x_2$ introduces an external magnetic field along the $x_1$ direction. The magnetic field in the $x_1$ direction breaks the rotational symmetry $SO(3)$ to the $SO(2)$. Therefore, the metric at finite temperature becomes magnetic field dependent and anisotropic which is similar to \cite{1802.05652}. We assume that \begin{equation}
g_{11}=e^{c_{1} C(B)z^{2}}, ~g_{22}=e^{c_{2} C(B) z^{2}},
\end{equation}
where the coefficients $c_{1,2}$ are two constants and $C(B)$ is an arbitrary function of $B$. Note that, at the boundary $z \to 0$, $b=g_{11}=g_{22}=1$.

The black hole entropy and temperature for this ansatz are well defined as
\begin{eqnarray}
s(z_h)&=&\frac{w_E^{3}g_{22}\sqrt{g_{11}}}{4}\bigg\vert_{z=z_h},\\
T(z_h)&=&-\frac{\partial_z b}{4\pi}\bigg\vert_{z=z_h} .
\end{eqnarray}

Before we start to solve the equations of motion, it is worth to verify the null energy condition (NEC) to guarantee the consistency of the gravitational model. The NEC can be expressed as
\begin{equation}
T_{\mu \nu}N^{\mu}N^{\nu} \geq 0, \label{NEC}
\end{equation}
where $T^{\mu\nu}$  is the energy-momentum tensor of the matter fields. The null vector $N^\mu$ satisfies the condition $g_{\mu\nu}N^{\mu}N^{\nu}=0$ and could be chosen as 
\begin{equation}
N^{\mu}=\frac{1}{\sqrt{b\left(  z\right)  }} N^{t}
+\frac{\sin\theta \cos\theta}{\sqrt{g_{11}}} N^{x_1}
+\frac{\sin^2\theta}{\sqrt{2 g_{22}}} \left( N^{x_2}+N^{x_3}\right)
+\cos\theta\sqrt{b\left(z\right)} N^{z},
\end{equation}
for arbitrary parameter $\theta$. Then the NEC Eq. (\ref{NEC}) becomes
\begin{equation}
\left( A_t'^2 + \frac{B^2\sin^2\theta}{e^{2c_2 C z^2} }\right) \frac{f \sin^2\theta}{2w_E^4}  + \frac{b \phi'^2 \cos^2\theta}{2w_E^2} \geq 0, \label{NEC constraint}
\end{equation}
which demands that the kinetic gauge function $f$ and $\phi'^2$ should be positive. (Which is the positive energy definite through the Legendre transformation of Lagrange density.)

Plugging the ansatz Eq. (\ref{eq_metric}) and Eq. (\ref{eq_ansatz}) into Eqs. (\ref{eq_eom_phi} - \ref{eq_eom_g}), the equations of motion reduce to
\begin{eqnarray} 
A_t''&+& \left( \frac{f'}{f} + \frac{w_E'}{w_E} + \left( c_1+2c_2\right) C z \right) A_t' =0\label{eom_D_At},\\
\phi'^2 &+& \frac{6w_E''}{w_E}-\frac{12w_E'^2}{w_E^2} 
+ 2( c_1+2c_2)C + 2( c_1^2+2c_2^2)C^2 z^2 =0 \label{eom_D_phi},\\ 
b''&+& \left( \frac{3w_E'}{w_E} + c_1 C z \right) b'
- 2 c_2 C \left( 1+(c_1+2c_2)C z^2+\frac{3z w_E'}{w_E} \right) b
- \frac{f}{w_E^2} \left( A_t'^2 + \frac{B^2}{e^{2c_2 C z^2} } \right) =0 \label{eom_D_b},\\
V &+& \frac{3 b w_E''}{w_E^3} + \frac{6 b w_E'^2}{w_E^4} + \left( 3 b C (c_1+4 c_2)z + \frac{9b'}{2} \right) \frac{w_E'}{w_E^3} \notag \\
&+&  \frac{f B^2 }{2 w_E^4 e^{2c_2 C z^2} } + \frac{b''+(c_1+6c_2) C z b'+4 c_2 C \left(1 + (c_1+2c_2) C z^2\right) b}{2w_E^2}=0 \label{eom_V},\\
\phi''&+& \left( \frac{3w_E'}{w_E}+ \frac{b'}{b} 
+ \left( c_1+2c_2\right)C z\right) \phi'
+ \left(A_t'^2 -\frac{B^2}{e^{2c_2 C z^2} } \right) \frac{ f'(\phi) }{2b w_E^2}
-\frac{w_E^2}{b} V'(\phi) =0. \label{eom_D_V}
\end{eqnarray}
Since there are four physical quantities {$A_t, \phi, b, V$} so we only need four equations of motion Eqs. (\ref{eom_D_At} - \ref{eom_V}). The last Eq. (\ref{eom_D_V}) is the redundancy one because of the Bianchi identity.

In order to solve the equations of motion, we need to consider boundary conditions for the first and second order ODEs. The asymptotic $AdS$ condition in the UV limit at the conformal boundary $z \to 0$, and the regular condition in the IR limit at the black hole horizon $z \to z_h$ are imposed.
\begin{enumerate}[(i)]
	\item $z \to 0:$
	\begin{eqnarray}
	&&d(0)=\phi(0)=0,~ b(0)=1 \label{bdy_0},\\
	&&A_t(0)=\mu.
	\end{eqnarray}
	\item $z=z_h:$
	\begin{equation}
	A_t(z_h)=b(z_h)=0, \label{bdy_zh}
	\end{equation}
\end{enumerate}
where $\mu$ is the chemical potential. According to the holographic dictionary of the gauge/gravity correspondence, 
\begin{equation}
A_t(z)=\mu-\rho z^2+...
\end{equation}
In addition, as the magnetic field shrinks to zero, the spacial symmetry should restore and return to the asymptotic $AdS$ near the boundary,
\begin{equation}
g_{11}=g_{22}=1.
\end{equation}

\subsection{Analytical Solution}
The equations of motion Eqs. (\ref{eom_D_At} - \ref{eom_D_V}) are very complicated. However, we found that the equations system can be analytically solved if we set $c_2=0$. With this choice, the equations reduces to
\begin{eqnarray} 
&&A_t''+ \left( \frac{f'}{f}+ \frac{w_E'}{w_E} +  c_1 C z \right) A_t'=0 \label{eq_sol_At},\\
&&\phi'^2 +\frac{6w_E''}{w_E}-\frac{12w_E'^2}{w_E^2} +  2c_1 C \left(1+ c_1C z^2\right)=0 \label{eq_sol_phi},\\ 
&&b''+ \left( \frac{3w_E'}{w_E} + c_1 C z \right) b'
-\frac{f}{w_E^2} \left( A_t'^2 + B^2\right) =0 \label{eq_sol_b}, \\
&&V+\frac{3 b w_E''}{w_E^3}+\frac{6 b w_E'^2}{w_E^4}+\left( 3 b c_1 C z +\frac{9b'}{2} \right) \frac{w_E'}{w_E^3}
+\frac{ b''+c_1  C z b'}{2w_E^2}+\frac{f B^2 }{2w_E^4 } =0. \label{eq_sol_V}
\end{eqnarray}
The fields $A_t,\phi'$ and $b$ can be analytically integrated from the Eqs. (\ref{eq_sol_At} - \ref{eq_sol_b}) as
\begin{eqnarray} 
&&A_{t}(z) = \mu \left[1-\frac{I_2(z)}{I_2(z_h)}\right], \label{At}\\
&&\phi'=\sqrt{-\frac{6w_E''}{w_E}+\frac{12w_E'^2}{w_E^2} - 2c_1 C \left(1+ c_1C z^2\right)}, \label{phi}\\ 
&&b(z) = 1-\dfrac{I_1(z)}{I_1(z_h)}+\frac{\mu^2}{I_2^2(z_h)I_1(z_h)}
\left\vert
\begin{array}
[c]{cc}%
I_1(z_h) & I_{12}(z_h)\\
I_1(z) & I_{12}(z)
\end{array}
\right\vert
+ \frac{B^2}{I_1(zh)}
\left\vert
\begin{array}
[c]{cc}%
I_1(z_h) & I_{13}(z_h)\\
I_1(z) & I_{13}(z)
\end{array}
\right\vert,\label{g-A}
\end{eqnarray}
with the help of the following integrals,
\begin{eqnarray}
I_1(z) &=& \int_{0}^{z} \frac{dy}{w_E^3 e^{\frac{1}{2}c_1Cy^2}}, \label{I1}\\
I_2(z) &=& \int_{0}^{z} \frac{dy}{w_E fe^{\frac{1}{2}c_1Cy^2}}, \label{I2}\\
I_3(z) &=& \int_{0}^{z} w_E fe^{\frac{1}{2}c_1Cy^2} dy, \label{I3}\\
I_{12}(z) &=& \int_{0}^{z} I'_1(y)I_2(y)dy=I_1(z)I_2(z)-\tilde{I}_{12}(z),\\
I_{13}(z) &=& \int_{0}^{z} I'_1(y)I_3(y)dy=I_1(z)I_3(z)-\tilde{I}_{13}(z). \label{I13}
\end{eqnarray}
It is apparent that the blackening factor $b(z)$ receives the contribution from both chemical potential $\mu$ and magnetic field $B$. In addition, if the magnetic field is turned off, the family of solutions will reduce to the original EMS system \cite{1301.0385}.

The potential $V(z)$ can be obtained from Eq. (\ref{eq_sol_V}). It is straightforward to reconstruct the potential $V(\phi)$ by $z$-expansion of both potential $V(z)$ and scalar field $\phi(z)$,
\begin{eqnarray}
V(\phi)&=&-12+\frac{(-3)}{2}\phi^2+...,
\end{eqnarray}
with the coefficient $-3$ being exactly the $m^2$ of the scalar field $\phi$, which satisfies the BF bound implying that the gravitational background is stable and $-12$ relates to the cosmological constant in $AdS_5$. This process is the so-called potential reconstruction.

To consider meson spectrum, we add a matter action of a probe vector field $V$ into the background,
\begin{equation}
S_{m} = -\frac{1}{16\pi G_{5}} \int d^{5}x\sqrt{-g}
{\frac{f\left(\phi\right)}{4}F_V^{2}}.
\end{equation}
The equation of motion for the vector field reads
\begin{equation}
\nabla_\mu [f(\phi)F_V^{\mu\nu}]=0.
\end{equation}
We use the gauge invariance to fix the gauge $V_z=0$ and write $V_i(\vec x,z)=\phi(\vec x)v_i(z)$with $\nabla^2 \phi(\vec x)=m^2\phi(\vec x)$. The equation of motion of the transverse vector field $V_\mu$ ($\partial^\mu V_\mu=0$) in the background Eq. (\ref{eq_metric}) is
\begin{equation}
v''_i+\left(\frac{b'}{b}+\frac{w_E'}{w_E}+\frac{f'}{f}-c_1 Cz\right)v'_i+\frac{m^2}{b}v_i=0,
\end{equation}
which can be brought to the Schr\"{o}dinger equation
\begin{equation}
\psi_i''+U(z)\psi+m^2 \psi_i=0,
\end{equation}
where the potential is
\begin{equation}
U(z) = \frac{X''}{X}-\frac{2X'^2}{X^2},
\end{equation}
with
\begin{equation}
X= \left(\frac{e^{c_1 Cz^2/2}}{bfw_E}\right)^{1/2}.
\end{equation}
To realize the linear Regge trajectories for the meson mass spectra, we choose the gauge kinetic function as 
\begin{equation}
f(z) =\frac{e^{-\left(R_{gg} + \frac{c_1 C}{2}\right)z^2  }}{zw_E} .
\end{equation}
At $T=\mu=B=0$, we thus have
\begin{equation}
U(z) = \frac{3}{4z^2}+R_{gg}^2 z^2.
\end{equation}
which leads to the linear mass spectrum $m_n^2=4R_{gg}n$, and the parameter $R_{gg}$ can be fitted by the Regge spectra, for instance the $J/\psi$ meson.

For the heavy quark sector, we choose the deformed factor $d(z)=\frac{-R_{gg}}{3}z^2-p z^4$ so that the warped factor $w_{E}(z) = \frac{1}{z}\exp\left({\frac{-R_{gg}}{3}z^2-pz^4}\right)$.  The parameter $p$ recalls the transition point at $\mu=B=0$ which can be fitted from lattice QCD. The parameter $R_{gg}$ and $p$ have been addressed in \cite{1301.0385,1703.09184} for the vanishing magnetic field with $R_{gg}=1.16$ and $p=0.273$.

With the above choice, the integrals Eqs. (\ref{I1} - \ref{I13}) are monotonously growing functions of $z$ from zero, and the integrals Eq. (\ref{I2}) and Eq. (\ref{I3}) do not depend on the magnetic field $B$.

To investigate the magnetic effects on the phase diagram, we need to explicitly fix the function $C(B)$ in the component of the metric $g_{11}=e^{c_{1} C(B)z^{2}}$. Dimension analysis \footnote{More precisely speaking that $\frac{T}{\sqrt{B}}$ should be dimensionless.} restricts  $C(B) \sim B$, we then take $C(B)=B$ without loss of generality. Subsequently, $c_1 \leq 0$ ensures the correct monotonous behavior of entropy as the size of BH changes. For simplicity, we set $c_1=-1$ in the later calculations.

\section{Phase Transition in QCD}
In this section, we study the phase structure for the black hole background which we obtained in the last section. The phase transition between the black holes correspond to both the confinement-deconfinement phase transition and the chiral phase transition in the dual holographic QCD theory \cite{1301.0385}.

In the presence of magnetic field, the QCD phase diagram is sensitive to the background magnetic field. It has been showed that the effect of background magnetic field on the transition temperature depends on the masses of the dynamical quarks. For light quarks with physical mass, lattice simulation reveals the IMC phenomenon for small chemical  potential \cite{1111.4956,1206.4205,1303.3972,1312.5628,1711.02884,1909.09547}. While for heavy quarks, the MC phenomenon is observed \cite{1005.5365,1203.3360,1209.0374}. In this work, we focus on the model of heavy quarks, since we have fixed the model parameters in terms of linear Regge behavior of $J/\psi$ meson and these parameters capturing the dynamics of the heavy quarks sector.

\subsection{Magnetic Effects on Temperature}
Using the solutions Eqs. (\ref{At} - \ref{g-A}), the BH entropy density and the Hawking temperature are straightforwardly calculated,
\begin{eqnarray}
s(z_h)&=&\frac{1}{4I'_1(z_h)},\\
T(z_h)&=& T_0
\left( 1 + \mu^{2} T_{\mu} + B^{2} T_{B} \right),\label{temperture}
\end{eqnarray}
where 
\begin{equation}
T_0=\frac{I'_1(z_h)}{4\pi I_1(z_h)},~ T_{\mu}=-\frac{\tilde{I}_{12}(z_h)}{I_2^2(z_h)},~ T_{B}=-\tilde{I}_{13}(z_h).
\end{equation}
From the explicit expression of the integral $I_1(z)$, we can show that
\begin{equation}
T_0=\frac{I'_1(z_h)}{4\pi I_1(z_h)}=\frac{1}{4\pi w_E^3(z_h) \int_{0}^{z_h} e^{-\frac{1}{2}B(z_h^2-y^2)}\frac{dy}{w_E^3(y) }}\ge \frac{1}{4\pi w_E^3(z_h) \int_{0}^{z_h} \frac{dy}{w_E^3(y) }}=\left. T_{0}\right\vert _{B=0} ,
\end{equation}
which implies that magnetic field will enhance the BH temperature $T(z_h)$ by $T_0$. On the other hand, both $T_\mu$ and $T_B$ are negative due to the positive integrals. The effect from $T_\mu$ and $T_B$ grows for larger magnetic $B$ and bigger horizon $z_h$, so that magnetic field will reduce the BH temperature eventually for large $B$. The two effects compete each other as we changing the magnetic field.
\begin{figure}[t!]
	\begin{center}
		\includegraphics[
		height=1.3in, width=1.7in]
		{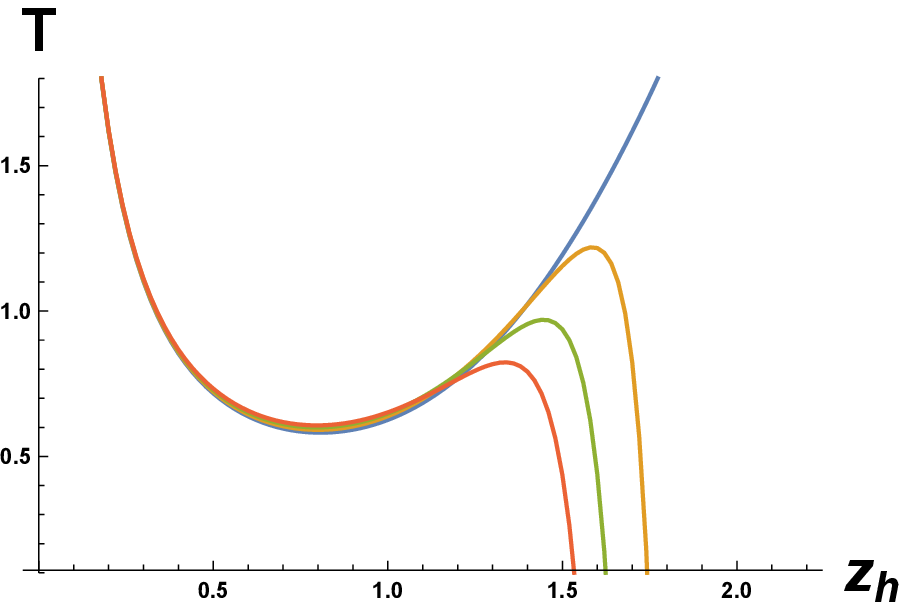}
		\includegraphics[
		height=1.3in, width=1.7in]
		{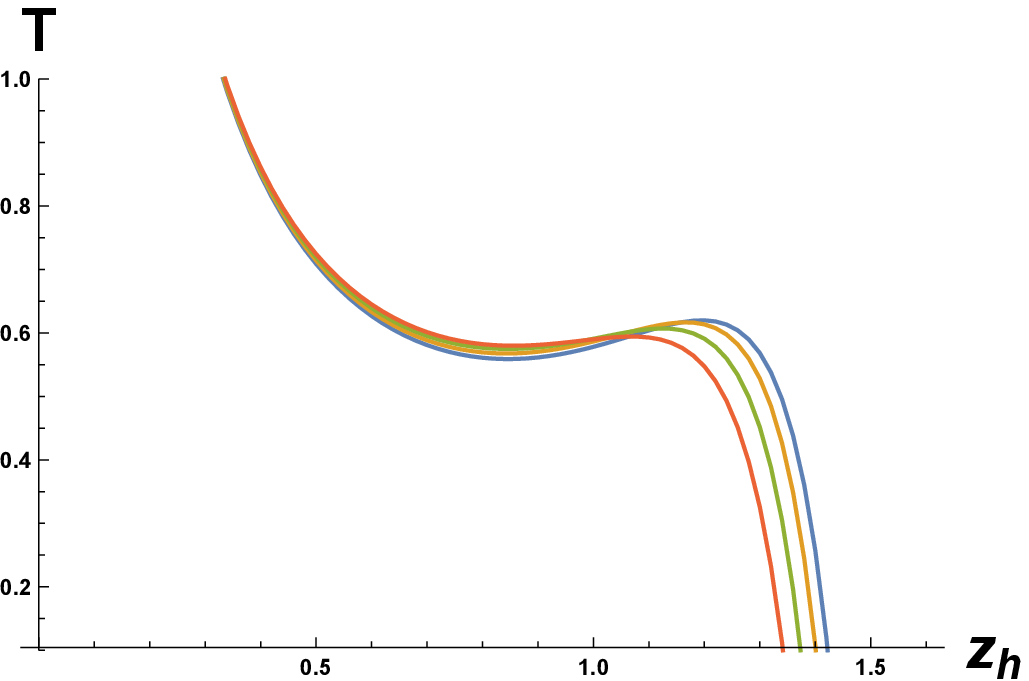}
		\includegraphics[
		height=1.3in, width=1.7in]
		{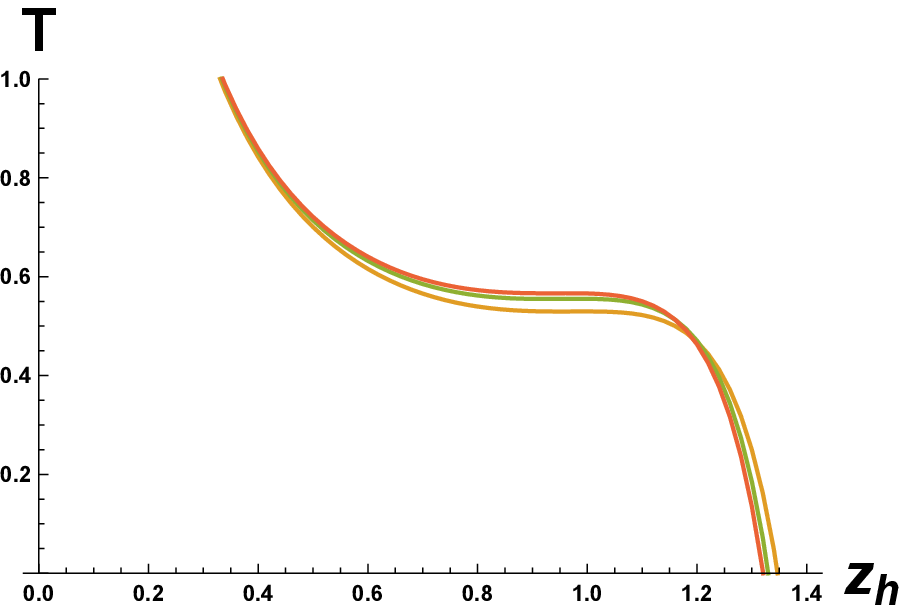}
		\includegraphics[
		height=1.3in, width=1.7in]
		{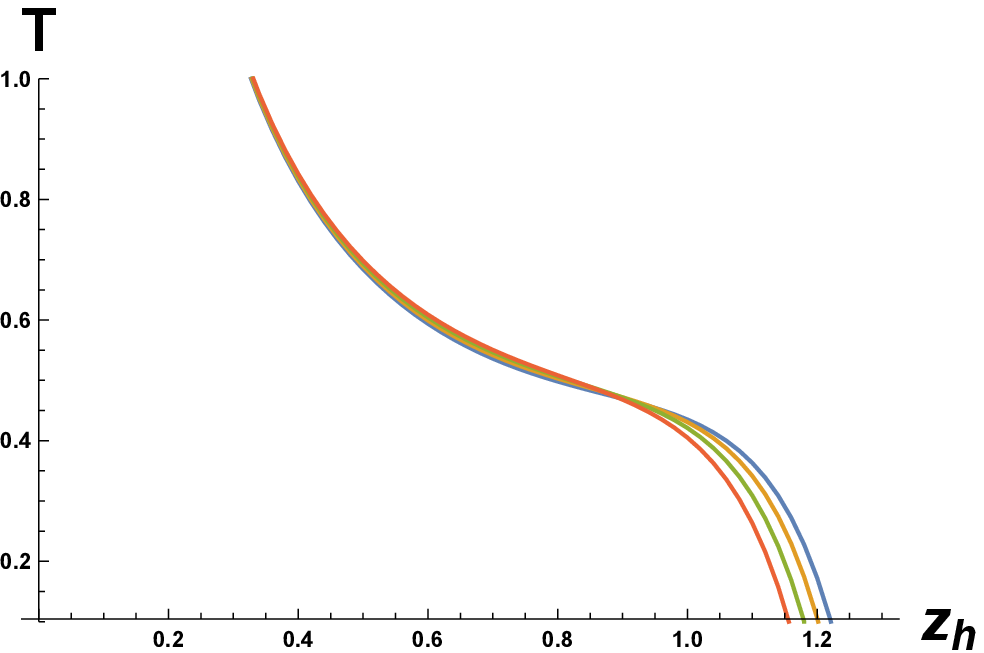}
	\end{center}
	\caption{Temperature vs horizon. From left to right: $\mu=0$, $\mu=0.5$, $\mu = \mu_{CEP}(B)$, $\mu = 1.0$. In each figure, from top to bottom: $B=0, 0.2, 0.4, 0.6$. Interestingly, the Hawking-Page transition only happens in $\mu=B=0$.} \label{fig_T_heavy}
\end{figure}

The black hole temperature in the presence of the background magnetic field $B$ for heavy quarks at different chemical potential are plotted in Fig.\ref{fig_T_heavy}. For small chemical potential $\mu<\mu_{CEP}$, the temperature is a multiple-valued function of the black hole horizon, which indicates that there would be a first-order phase transition between the large and the small sizes of black holes. For $\mu>\mu_{CEP}$, the temperature becomes monotonous and the phase transition could reduces to a crossover. At the critical point $\mu=\mu_{CEP}$, we expect a second-order phase transition.

The influence of magnetic field on temperature can be read from Fig.\ref{fig_T_heavy}. The temperature is enhanced at small horizon $z_h$ due to the effect from $T_0$, while is reduced at large $z_h$ due to the effect from $T_\mu$ and $T_B$. To investigate the magnetic effect in detail, we take the derivative of the temperature with magnetic field,
\begin{eqnarray}
\frac{d T}{d B} 
&=&\frac{I_{1}^{\prime }}{8\pi I_{1}}\left( H(z_h) -\frac{\mu ^{2}}{%
	I_{2}^{2}}\int_{0}^{z_h}\left[ H(z_h)+\frac{I_{1y}\left( y\right) }{I_{1}\left( y\right) }\right]
I_{1}\left( y\right) I_{2}^{\prime }\left( y\right) dy\right) \notag \\
& &-\frac{B^{2}I_{1}^{\prime }}{8\pi I_{1}}\int_{0}^{z_h}%
\left[ H(z_h)+\frac{%
	I_{1y}\left( y\right) }{I_{1}\left( y\right) }\right] I_{1}\left( y\right)
I_{3}^{\prime }\left( y\right) dy-\frac{BI_{1}^{\prime }\tilde{I}_{13}}{2\pi I_{1}},
\label{dT/dB}
\end{eqnarray}
where
\begin{eqnarray}
H(z_h) &=& z_h^{2}-\frac{I_{1y}\left( z_h\right) }{I_{1}\left( z_h\right)} \text{ and }  I_{1y}\left( z_h\right)=\int_{0}^{z_h}y^{2}I_{1}^{\prime }dy .
\end{eqnarray}
It is easy to show that $H(z_h)\ge 0$ by using the monotonicity of the integral $I_1(z)$. 

For small magnetic field, we can neglect the last term in Eq. (\ref{dT/dB}),
\begin{eqnarray}
\frac{dT}{dB} 
&\simeq&\frac{I_{1}^{\prime }}{8\pi I_{1}}\left( H(z_h) -\frac{\mu ^{2}}{%
	I_{2}^{2}}\int_{0}^{z_h}\left[ H(z_h)+\frac{I_{1y}\left( y\right) }{I_{1}\left( y\right) }\right]
I_{1}\left( y\right) I_{2}^{\prime }\left( y\right) dy\right),
\end{eqnarray}
which is positive for not too large chemical potential $\mu$ since $I_1>0$. This implies the MC behavior for small chemical potential.

On the other hand, for large magnetic field $B$, only the last term in Eq. (\ref{dT/dB}) is important,
\begin{eqnarray}
\frac{dT}{dB} 
&\simeq&-\frac{B^{2}I_{1}^{\prime }}{8\pi I_{1}}\int_{0}^{z_h}%
\left[ H(z_h)+\frac{%
	I_{1y}\left( y\right) }{I_{1}\left( y\right) }\right] I_{1}\left( y\right)
I_{3}^{\prime }\left( y\right) dy-\frac{BI_{1}^{\prime }\tilde{I}_{13}}{2\pi I_{1}}<0,
\end{eqnarray}
which implies the IMC behavior for large magnetic field.

\subsection{Magnetic Effects on QCD Phase Diagram}
As we argued in the previous section, the rough structure of QCD phase transition can be read from the behavior of black hole temperature. However, to obtain the exact phase diagram, we need to calculate the free energy of the thermodynamic system. The free energy in grand canonical ensemble can be obtained from the first law of thermodynamics,
\begin{equation}
F=\epsilon-Ts-\mu\rho-MB, \label{eq_def_F}
\end{equation}
where $\rho=\dfrac{\mu}{2 I_2(z_h)}$ is the baryon density, $M$ represents the magnetization which is associated to $B$, and $\epsilon$ labels the internal energy density. Comparing the free energies of BHs at the same temperature for certain finite value of chemical potential, we are able to obtain the phase structure of BHs which gives the phase diagram of the holographic QCD due to AdS/CFT correspondence. At fixed volume, the differential of the free energy is defined as \cite{1909.09547}
\begin{equation}
dF=-sdT-\rho d\mu-MdB.
\end{equation}
For the fixed chemical potential $\mu$ and magnetic field $B$, the free energy can be evaluated by the following integral,
\begin{equation}\label{freeenergy}
F= -\int sdT=\int_{z_h}^\infty s(\bar{z_h}) T'(\bar{z_h}) d \bar{z_h}
\end{equation}
where we have normalized the free energy to vanish at $z_h \to \infty$.

\begin{figure}[t!]
	\begin{center}
		\includegraphics[
		height=1.4in, width=2.3in]
		{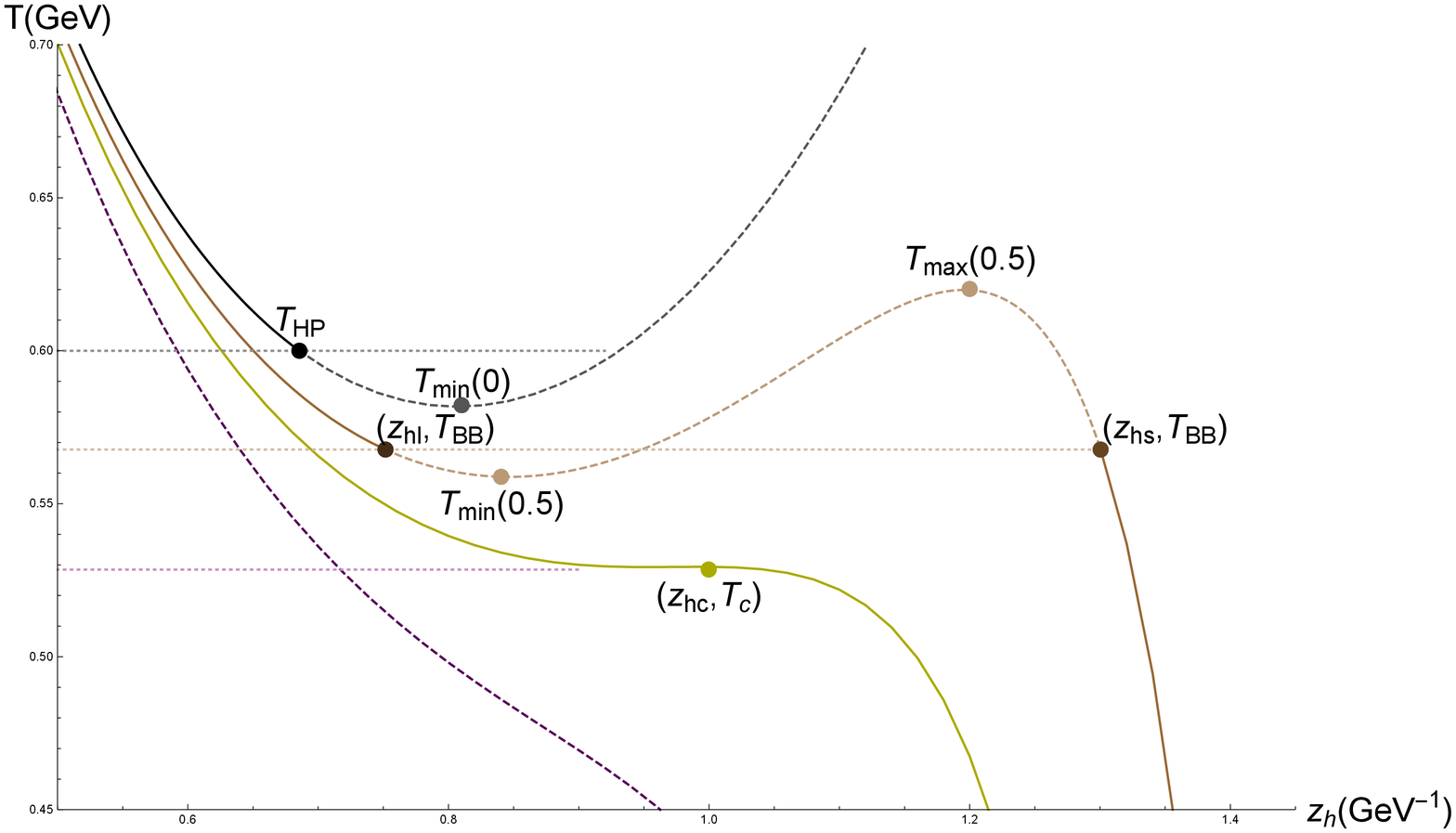}
		\includegraphics[
		height=1.4in, width=2.3in]
		{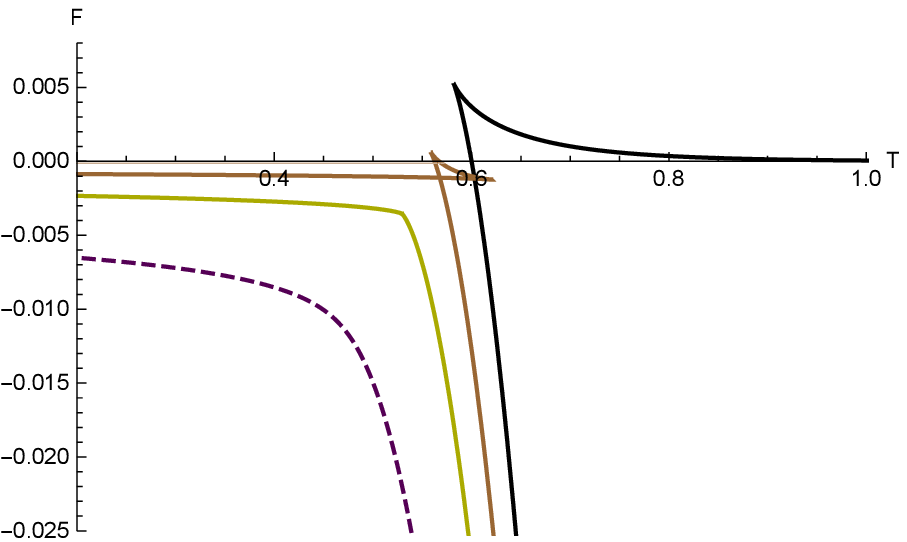}
		\includegraphics[
		height=1.4in, width=2.3in]
		{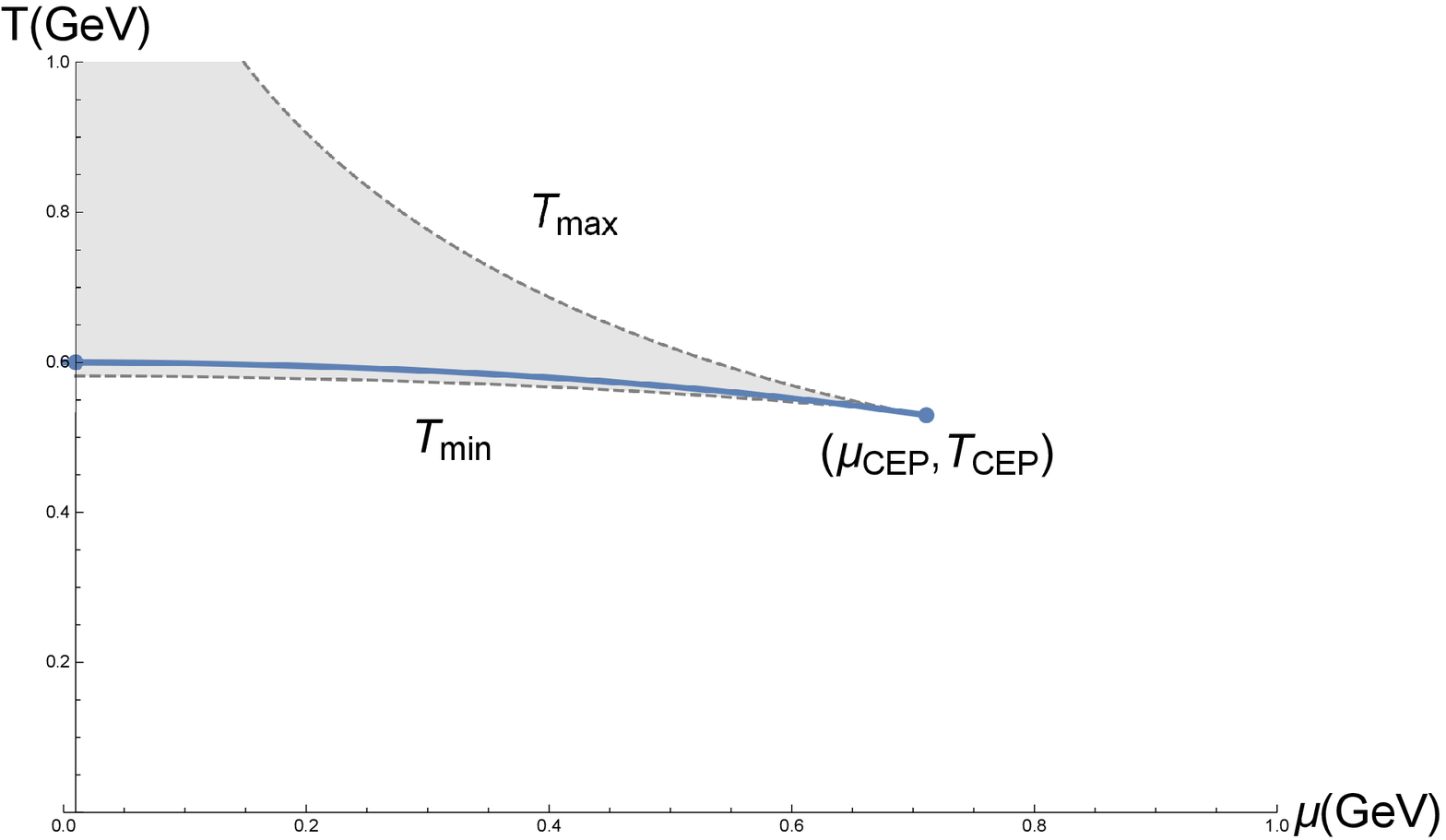}
	\end{center}
	\caption{(Left) Temperatures vs horizon. From top to bottom: $\mu=0, 0.5, \mu_{CEP}, 1.0$. (Middle) Free energies vs temperature for the corresponding $\mu's$. The intersection of the swallow-type patterns labels the first order phase transition. (Right) QCD phase diagram for heavy quarks. The phase transition line is within the shadow region which is enclosed by the curves of the local minimum and maximum temperatures that shrink to the CEP at $(\mu_{CEP},T_{CEP})=(0.714, 0.528).$}\label{fig_TB0}
\end{figure}

The thermodynamic properties for heavy quarks without magnetic field has been studied in \cite{1301.0385}. The temperature vs horizon is plotted in the left figure of Fig.\ref{fig_TB0}. At $\mu=0$, the temperature has a minimum value which implies the Hawking-Page phase transition. For a finite but small $\mu$, the temperature drops to zero at certain horizon, but a first order phase transition still happens at a finite temperature between the local minimum and maximum temperatures. From the local minimum and maximum for each chemical potential $\mu<\mu_{CEP
}$, we can partition a region where the phase transition could happen. The CEP of the phase diagram will take place at the specific temperature $T_{CEP}$ where the local minimum and maximum temperature are degenerated at characteristic chemical potential $\mu_{CEP}$. When the chemical potential is beyond a critical value $\mu>\mu_{CEP}$, the temperature becomes monotonous and the phase transition reduces to a crossover. The exact phase transition temperature can be obtained from the free energy. The free energy vs temperature is demonstrated in the middle figure of Fig.\ref{fig_TB0}. The intersection of the swallow-type patterns labels the first order phase transition temperature at each chemical potential where the free energy and temperature are equal at different horizons. As the chemical potential increasing, the swallow-type gradually compresses and eventually becomes a singular point where character the second order phase transition. When $\mu>\mu_{CEP}$, the free energy becomes smooth function with respect to temperature that implies the phase transition is weaken to a crossover. The phase diagram is delivered in the right figure of Fig.\ref{fig_TB0}. The shadow region is enclosed by the curves of the local minimum and maximum temperatures, which shrink to the CEP. The phase transition line is within the shadow region. We should remark that the behavior of the heavy quarks is in contrast to that of the light quarks, in which crossover occurs at small chemical potential and becomes phase transition for $\mu>\mu_{CEP}$.

\begin{figure}[t!]
	\begin{center}
		\includegraphics[
		height=1.4in, width=2.2in]
		{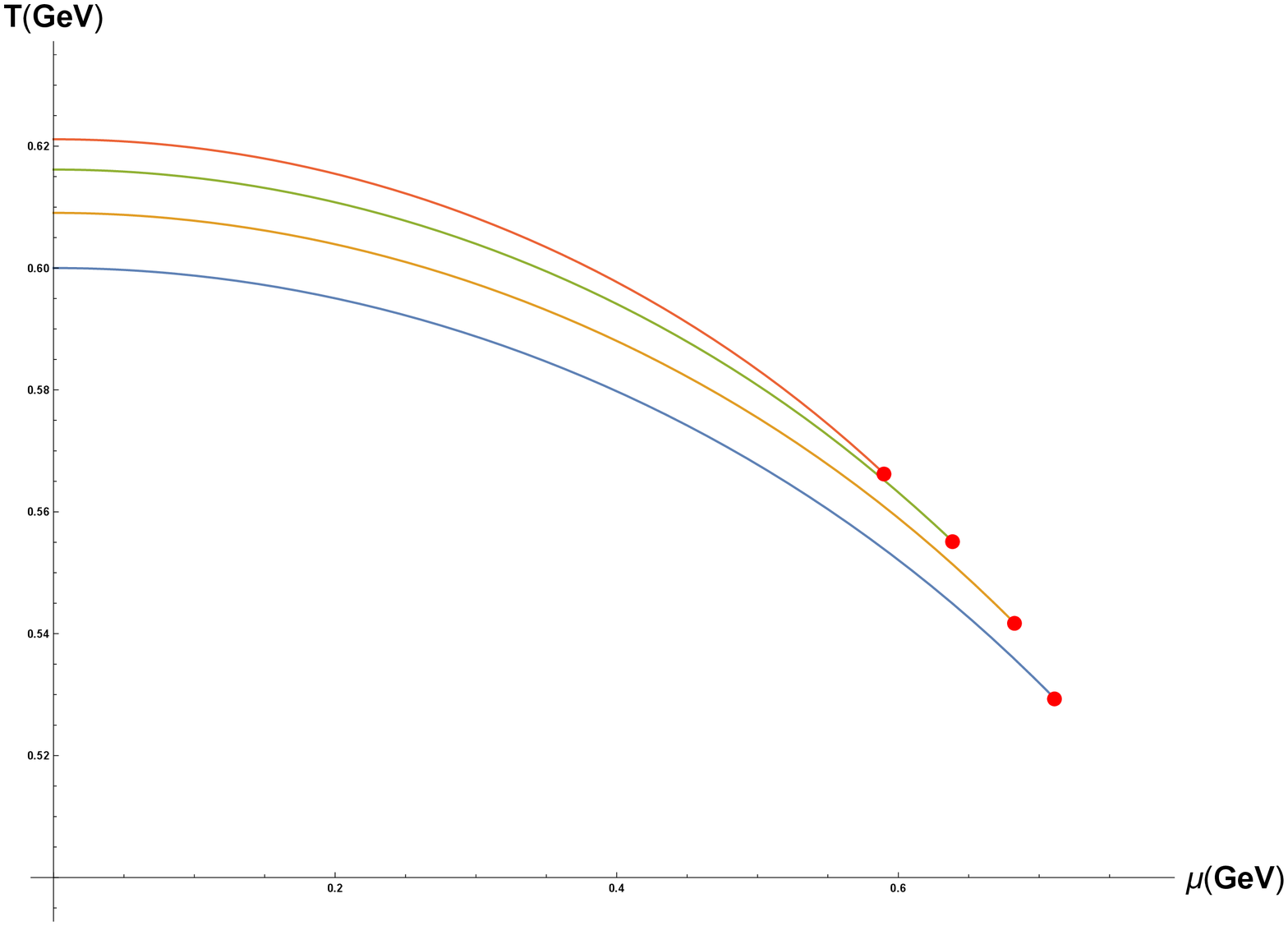}
		\includegraphics[
		height=1.4in, width=2.1in]
		{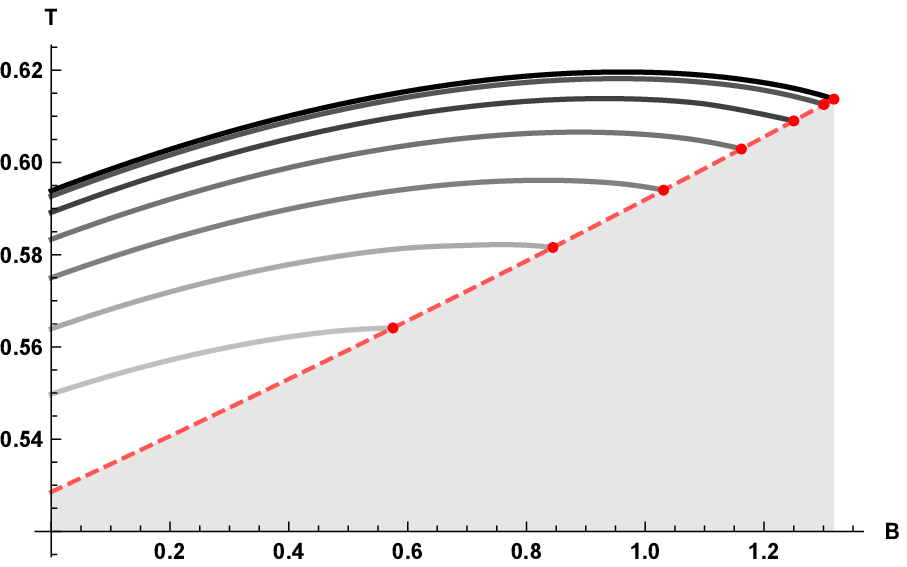}
		\includegraphics[
		height=1.6in, width=2.6in]
		{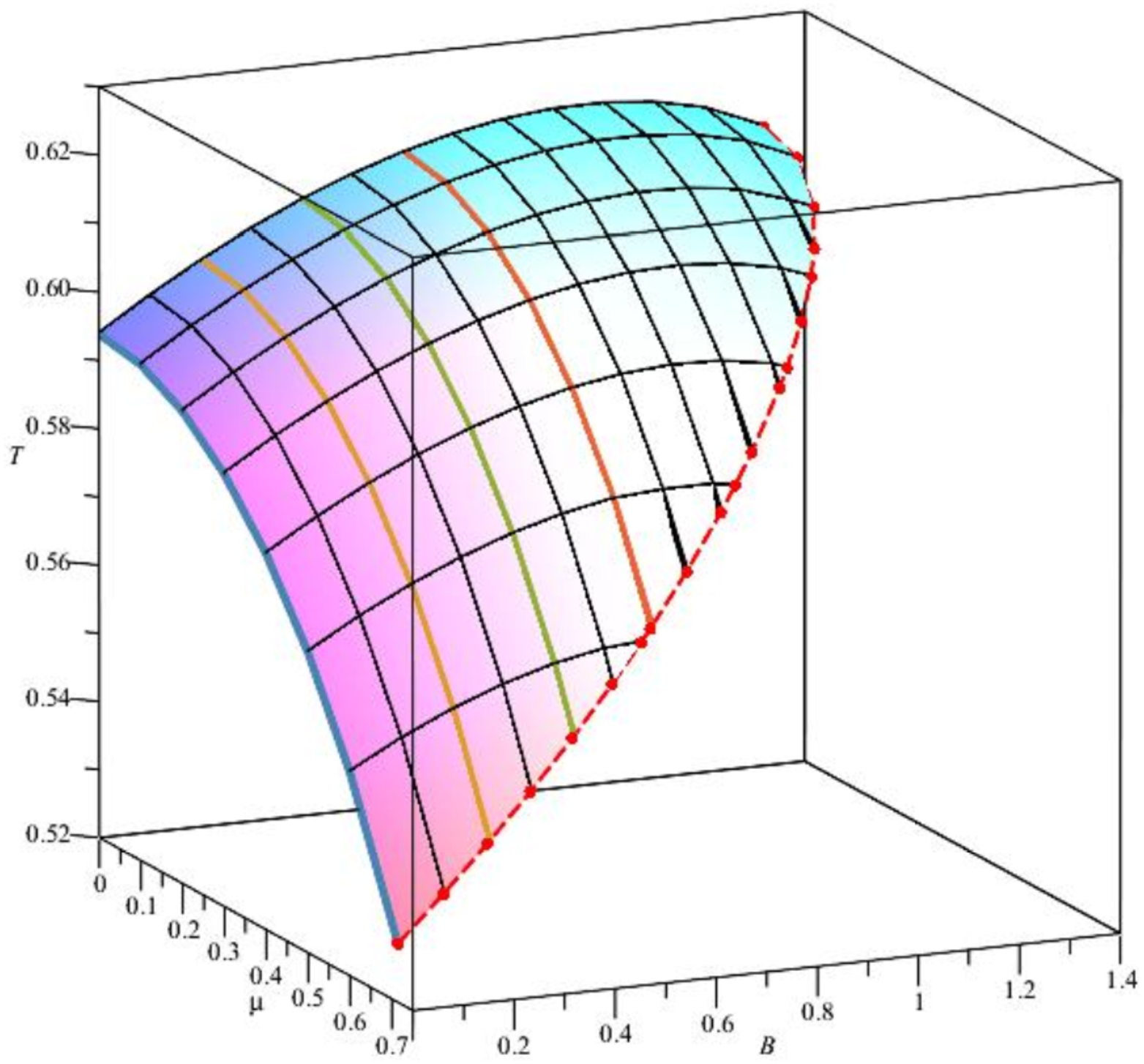}
	\end{center}
	\caption{(Left) Phase diagram on $T-\mu$ plane for $B=0, 0.2, 0.4, 0.6$ from bottom to top. The red dots label the CEPs for different magnetic fields. (Middle) Phase diagram on $T-B$ plane for $\mu=0, 0.1, 0.2, 0.3, 0.4, 0.5, 0.6$ from top to bottom. The red dots label the CEPs for different chemical potentials which are collected in the red dashed line with the shadow region standing for the crossover zone. (Right) The 3-dimensional phase diagram in $T-\mu-B$ axes. The curved surface represents the first order phase transition area and terminates at the red dashed boundary which is the collection of CEPs.} \label{fig_phaseTmuB_heavy}
\end{figure}

In the presence of magnetic field $B$, the behavior of free energy are qualitatively the same as that with zero magnetic field as showed in the middle figure of Fig.\ref{fig_TB0}. The exact values of the phase transition temperature changes with both chemical potential $\mu$ and  magnetic field $B$ are plotted in Fig.\ref{fig_phaseTmuB_heavy}.

For fixed $B$ fields, phase transition extends from $\mu=0$ to finite chemical potentials and terminates at a CEP, the red dots, as showed in the left figure of Fig.\ref{fig_phaseTmuB_heavy}. Increasing the magnetic field from zero enhances the phase transition temperature. This confirms the MC behavior as we expected from the behavior of the temperature. The MC phenomenon we found for the heavy quarks is consistent with the recent lattice simulations \cite{1808.07008,1904.10296}. Furthermore, it shows that the CEP of the phase transition moves towards to the lower chemical potential when the magnetic field increases, that is consistent with the recent works using PNJL model \cite{1509.01181,1712.08378,1712.08384}.

We plot the phase transition temperature vs magnetic field for different chemical potentials in the middle figure of Fig.\ref{fig_phaseTmuB_heavy}. For small magnetic field, it clearly shows that the phase transition temperature increases along the magnetic field, i.e. MC phenomenon as we have discussed. On the other hand, when the magnetic field $B$ become large enough, we observe that the phase transition temperature reduces with the increasing magnetic field. This justifies our conclusion of IMC behavior at large magnetic field by the argument of the competition between the contributions from $T_0$ and $T_B$ to the temperature in Eq. (\ref{temperture}). In our case, the flipping magnetic field $B \sim 0.96$ at $\mu=0$, which magnitude is very close to \cite{1111.4956,1501.03262}. Since the sign problem would not encounter at $\mu=0$, we expect this appearance can be attended in lattice QCD. We also notice that the phase transition will reach CEP before it turns to IMC from MC if the chemical potential is too large $\mu\gtrsim 0.6~GeV$. Another words, IMC only appears at relatively low baryon density, which statement has also been declared in \cite{1707.00872}. For each chemical potential, there exists a critical magnetic field $B_{CEP}(\mu)$ beyond that the phase transition becomes crossover. The CEPs for different chemical potentials are along the red dashed line with the shadow region standing for the crossover zone.

The full phase transition structure including both chemical potential and magnetic field is combined in a 3-dimensional phase diagram as plotted in the right figure of Fig.\ref{fig_phaseTmuB_heavy}. The left two diagrams are the 3-dimensional phase diagram projected on the $T-\mu$ and $T-B$ planes, respectively. The four lines with the fixed magnetic fields in the $T-\mu$ diagram are put with the corresponding colors in the 3-dimensional phase diagram. In addition, the red dashed line in the 3-dimensional phase diagram indicates the 3-dimensional CEP boundary, beyond which the phase transition becomes crossover.

\subsection{Critical End Point}
It is crucial to locate the CEP of the phase transition \cite{0106002,0210284,1712.08378,1908.02000}. It has been showed that the QCD phase transition temperature gradually cools down as QCD matter being more and more dense, i.e. increasing chemical potential. The CEP of the first order phase transition can be evaluated by the free energy Eq. (\ref{freeenergy}). 

\begin{figure}[t!]
	\begin{center}
		\includegraphics[
		height=1.5in, width=2.3in]
		{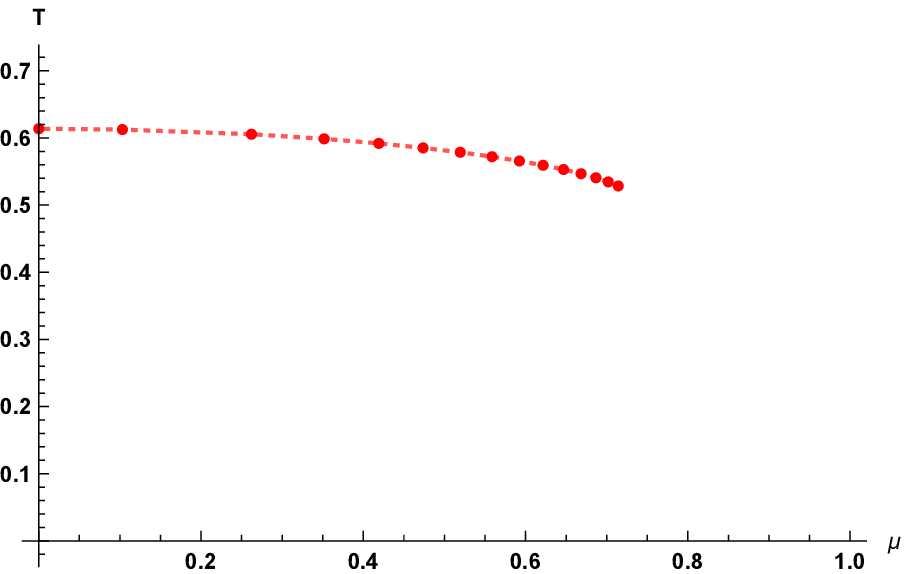}
		\includegraphics[
		height=1.5in, width=2.2in]
		{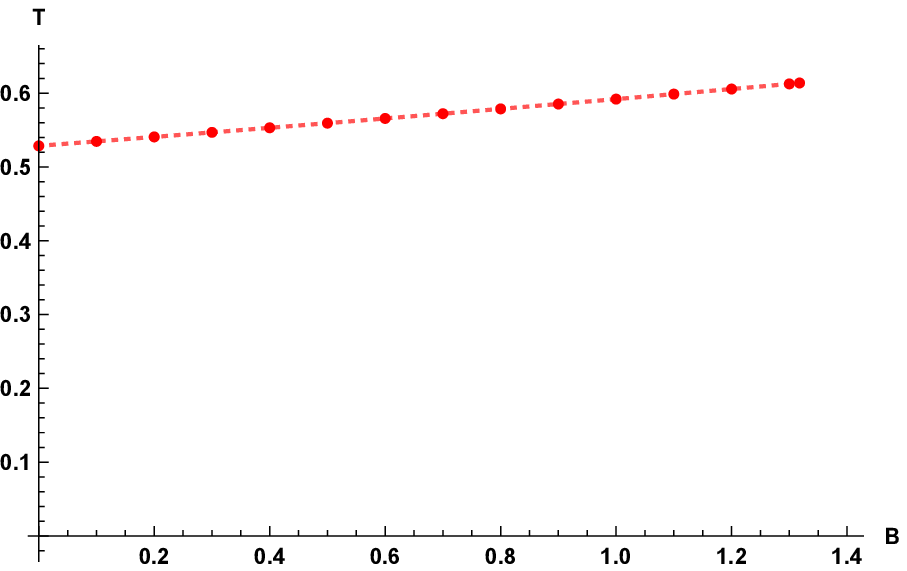}
		\includegraphics[
		height=1.5in, width=2.2in]
		{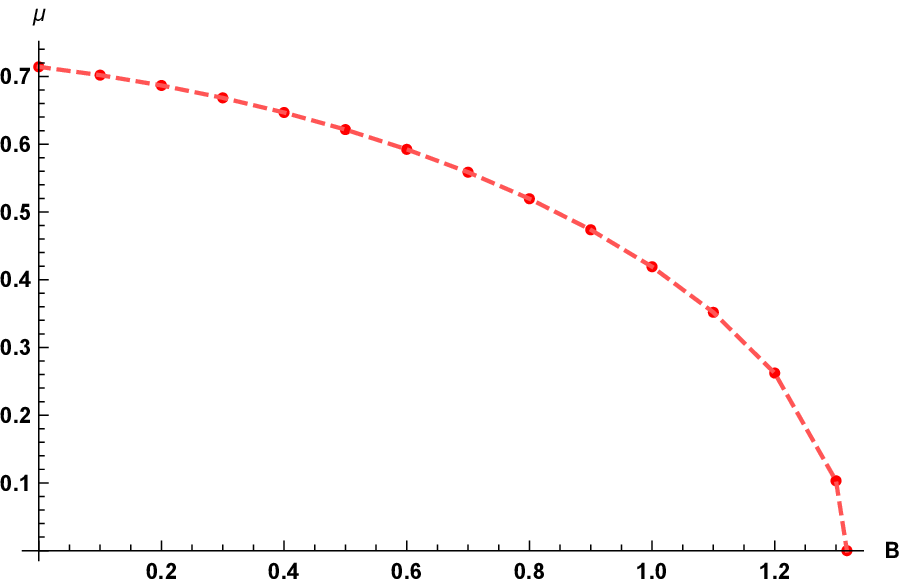}
	\end{center}
	\caption{The CEPs on $T-\mu$, $T-B$ and $\mu-B$ planes. We can observe that the CEP temperature are linear with the magnetic field. On the other hand, as magnetic field increasing, the chemical potential gradually decreases and eventually vanishes at $B \sim 1.31~GeV^2$ which implies that the whole transition line is weaken to a transition point. We can demonstrate this tendency in the $T-\mu$ plane. As the imposing magnetic field become stronger, the CEP temperature become higher but CEP baryon density become looser. Another words, the first order phase transition line become shorter, the crossover region become wider, as the magnetic field is stronger.} \label{fig_cepTB_heavy}
\end{figure}
The authors of \cite{1509.01181,1712.08378,1712.08384} have studied confinement-deconfinement phase transition by using PNJL model. They showed that the CEP moves towards lower chemical potentials with increasing magnetic field if considering a magnetic field dependent coupling $G(B)$, and the CEP could eventually approaches to the zero chemical potential for large enough magnetic field. Their result is only for light quarks, here we show that the similar behavior preserves for heavy quarks.

The tendency of CEP with magnetic field is plotted in the left figure of Fig.\ref{fig_cepTB_heavy}. When the magnetic field increasing, the CEP moves to the lower chemical potential and approaches to $T\sim 0.613 ~GeV$ at $\mu=0$. The CEP on $T-B$ plane is plotted in the middle figure of Fig.\ref{fig_cepTB_heavy}. When the chemical potential increasing, the CEP moves to the smaller magnetic filed and returns to the original CEP at $B=0$. It is interesting to observe that the CEP temperature are linear with the magnetic field. Furthermore, the CEP forms a closed boundary on $\mu-B$ plane as plotted in the right figure of Fig.\ref{fig_cepTB_heavy}. Beyond the boundary the phase transition becomes crossover.

\section{Conclusion}
In this paper, we have constructed an analytical holographic QCD model for heavy quarks in the presence of the external magnetic field by using the Einstein-Maxwell-Scalar system with potential reconstruction approach. We introduced both electrical and magnetic fields in a single bulk $U(1)$ gauge field. The electrical component is interpreted as the global baryon number conservation symmetry on the boundary theory by holography. While the magnetic component corresponds to the boundary external magnetic field. In this holographic setup, we have checked the NEC to ensure that the gravitational background is a stable. In addition, in this background, we have also considered the linear Regge spectrum of $J/\psi$ mesons. Since the parameters are fixed by the heavy mesons spectrum, the holographic model captures the characters of the heavy quarks sector.

We calculated the free energy to obtain the phase transition temperatures for heavy quarks at different chemical potentials and magnetic fields in the holographic QCD model in the anisotropic background. For small magnetic field, we found MC phenomenon that is consistent with the recent lattice results \cite{1808.07008,1904.10296}. While for large enough magnetic field, we found IMC phenomenon due to the competition between the contributions from $T_0$ and $T_B$ to the temperature in Eq. (\ref{temperture}). Thus the phase transition will change from MC to IMC as the magnetic field growing. However, we noticed that the phase transition will reach its critical end point before it turns to IMC from MC if the chemical potential is large enough $\mu\gtrsim~0.6GeV$. See Fig.\ref{fig_phaseTmuB_heavy} for the phase diagrams at different chemical potentials and magnetic fields.

In addition, there exists a extreme point for the phase transition for either chemical potential or magnetic field, i.e. the CEP. Beyond CEP, the phase transition becomes to crossover. We discover that, for fixed magnetic fields, the CEP moves to the lower chemical potential and eventually approaches to $\mu=0$. While for fixed chemical potentials, the CEP decreases with the magnetic field growing. It is interesting to observe that the CEP temperature are linear with the magnetic field. We do not understand the reason of this linear behavior and it is deserved to study in the future. See Fig.\ref{fig_cepTB_heavy} for the CEP at different chemical potentials and magnetic fields.

In this work, we only focused on the thermodynamics of the black hole to obtain the phase transition between two black holes with different sizes, by calculating their free energies in the hQCD model. It is also interesting to investigate the definite order parameters to explore further information associated with the phase transitions, such as confinement and chiral symmetry breaking. The two kinds of phase transitions will show different characteristic behaviors with respect to the magnetic field. We would like to study these corresponding order parameters to confirm such phenomenon in terms of holographic approach. We leave this part in the future.
\subsection*{Acknowledgements}
We would like to thank Umut Gürsoy, Danning Li, Xiaoning Wu, Lang Yu for useful discussion. S.H. also would like to appreciate the financial support from Jilin University and Max Planck Partner group. This work of Y.Y is supported by the Ministry of Science and Technology (MOST 106-2112-M-009 -005 -MY3) and National Center for Theoretical Science, Taiwan. The work of P.H.Y. was supported by the University of Chinese Academy of Sciences.

\end{document}